\documentclass[]{spie}  

\usepackage{amsmath,amsfonts,amssymb}
\usepackage{graphicx}
\usepackage{multirow}
\usepackage[colorlinks=true, allcolors=blue]{hyperref}

\title{Exoplanet Imaging Data Challenge, phase II: Characterization of exoplanet signals in high-contrast images}
\author[a]{Cantalloube~F.}
\author[b]{Christiaens~V.}
\author[b,c]{Cantero~C.}
\author[d]{Nasedkin~E.}
\author[c]{Cioppa~A.}
\author[b]{Absil~O.}
\author[e]{Bonse~J.~M.}
\author[f]{Delorme~P.}
\author[g]{Gomez-Gonzalez~C.}
\author[b]{Juillard~S.}
\author[h]{Mazoyer~J.}
\author[d]{Samland~M.}
\author[i]{Ruffio~J.-B.}
\author[c]{Van~Droogenbroeck~M.}

\affil[a]{Aix Marseille Univ, CNRS, CNES, LAM, Marseille, France}
\affil[b]{Space Sciences, Technologies \& Astrophysics Research (STAR) Institute, Universit\'e de Li\`ege, All\'ee du Six Ao\^{u}t 19c, B-4000 Li\`ege, Belgium}
\affil[c]{Montefiore Institute, Universit\'e de Li\`ege, 4000 Li\`ege, Belgium}
\affil[d]{Max-Planck-Institut f\"{u}r Astronomie, K\"{o}nigstuhl 17, Heidelberg 69117, Germany}
\affil[e]{Institute for Particle Physics and Astrophysics, ETH Zurich, Wolfgang-Pauli-Strasse 27, 8093, Zurich, Switzerland}
\affil[f]{Univ. Grenoble Alpes, CNRS, IPAG, F-38000 Grenoble, France}
\affil[g]{Barcelona Supercomputing Center, Spain}
\affil[h]{LESIA, Observatoire de Paris, Universit\'e PSL, CNRS, Sorbonne Universit\'e, Universit\'e de Paris, 5 place Jules Janssen, F-92195 Meudon, France}
\affil[i]{Department of Astronomy, California Institute of Technology, Pasadena, CA 91125, USA}

\authorinfo{Further author information: (Send correspondence to F.C.)\\ 
F.C.: E-mail: faustine.cantalloube@lam.fr}

\begin{document} 
\maketitle

\begin{abstract}
Today, there exists a wide variety of algorithms dedicated to high-contrast imaging, especially for the detection and characterisation of exoplanet signals. These algorithms are tailored to address the very high contrast between the exoplanet signal(s), which can be more than two orders of magnitude fainter than the bright starlight residuals in coronagraphic images. The starlight residuals are inhomogeneously distributed and follow various timescales that depend on the observing conditions and on the target star brightness. Disentangling the exoplanet signals within the starlight residuals is therefore challenging, and new post-processing algorithms are striving to achieve more accurate astrophysical results. 
The Exoplanet Imaging Data Challenge is a community-wide effort to develop, compare and evaluate algorithms using a set of benchmark high-contrast imaging datasets. After a first phase ran in 2020 and focused on the detection capabilities of existing algorithms, the focus of this ongoing second phase is to compare the characterisation capabilities of state-of-the-art techniques. The characterisation of planetary companions is two-fold: the astrometry (estimated position with respect to the host star) and spectrophotometry (estimated contrast with respect to the host star, as a function of wavelength). 
The goal of this second phase is to offer a platform for the community to benchmark techniques in a fair, homogeneous and robust way, and to foster collaborations.
\end{abstract}


\section{INTRODUCTION}
\label{sec:intro}  
High-contrast imaging (HCI) is a very demanding observation technique that requires dedicated state-of-the-art instrumentation in the visible to infrared on the largest ground-based telescopes. Even if other direct exoplanet detection/characterization techniques come into being, high-contrast imaging remains a technique of choice to obtain images of extrasolar systems, for which many archival data are publicly available to the community. When a point source is detected within high-contrast images, it primarily needs to be characterized, that is to say to estimate its relative projected \emph{position} and \emph{brightness} with respect to the target star (for one or more wavelengths). This characterisation step is essential to confirm that the detected signal is gravitationally bound to the star and is of a (sub)stellar nature, making it a firm companion detection. The characterization step is also fundamental to deduce the orbital and atmospheric properties of the companion, which allows to trace its formation history. In particular, the extracted spectrum, allows one to estimate the chemical composition, cloud properties, formation and migration histories, and in the future may allow the detection of biosignatures in the spectrum of rocky exoplanets\cite{madhusudhan2019atm,molliere2022atm}.

Post-processing techniques are an essential part of HCI whose goal is to gain several additional orders of magnitude in contrast in the images delivered by high-contrast instruments. These images are heavily contaminated by bright starlight residuals that follow various timescales and have an inhomogeneous spatial distribution. Indeed, typical observing sequences last about 1~hour during which observing conditions are likely to change and internal aberrations to evolve. In that context, extracting accurate information about the position and flux of an exoplanet signal that is more than one order of magnitude fainter than the starlight residuals is arduous. 

Therefore, exoplanet characterization in HCI is crucial both on the technical and astrophysical point of view. In this context, the goal of the second phase of the \emph{Exoplanet Imaging Data Challenge} (EIDC) is to focus on the characterization capabilities of most post-processing techniques dedicated to HCI that exist to date. Within this phase of the EIDC, we provide the community with four types of resources 
that are described in this paper: (1) an ensemble of HCI datasets with injected planetary signals, (2) a platform for participants to submit their results, (3) an assortment of metrics to evaluate and compare the submitted results, and (4) various software tools to help the HCI community to make the most out of this challenge. The main objective of this second phase of the EIDC is to get insights about the characterizations accuracy of each algorithm depending on the provided datasets. This way the EIDC will (i) provide guidance on the most appropriate techniques to be used by observers, depending on the scientific context, (ii) identify potential systematic errors in these techniques, (iii) clear new exploration pathways for improving our current tools and (iv) offer a set of data and evaluation metrics to assess the performance of the next generation of post-processing algorithms that can be used for publications.

The second phase of the EIDC has been launched on the 25th of April 2022, and the closing date has been postponed to 31st of December 2022. The main information about the \emph{Exoplanet Imaging Data Challenge} initiative can be found on the dedicated website\footnote{\url{https://exoplanet-imaging-challenge.github.io/}}, which contains two tabs dedicated to the first and second phases (top-right drop-down menu). The ensemble of data sets for this characterization phase are hosted on a Zenodo\footnote{\url{https://zenodo.org/record/6902628}} open-access repository. The participants are invited to submit their results via the EvalAI\footnote{\url{https://eval.ai/web/challenges/challenge-page/1717/}} competition platform, which computes a basic evaluation metrics that we previously defined to publicly display the obtained score on a leaderboard. 

In the following, we first describe the data sets provided to the participants, containing $21$ injected planetary signals, that are available on the Zenodo repository (Sect.~\ref{sec:data}). We then describe the evaluation procedure used for the leaderboard ranking on the evalAI platform for this characterization phase (Sect.~\ref{sec:eval}). We then give two examples of retrieval performed by two traditional post-processing algorithms, ANDROMEDA and PCA-NEGFC (Sect.~\ref{sec:preres}).
At last, we propose a path-forward for the next phases of the EIDC and for leveraging the limitations noted during the first and second phases. The co-authors of the present paper are all part of the working group preparing this second phase of the EIDC and/or kindly provided  with the pre-reduced HCI datasets used in this second phase (contact: \url{exoimg.datachallenge@gmail.com}). 
After the deadline, this working group will work on a complete publication including the results from all the submissions provided.

\section{Data sets}
\label{sec:data}
For this characterization challenge (EIDC, phase 2), we gathered 8 data sets from two of the latest generation of high-contrast low-resolution integral field spectrographs: VLT/SPHERE-IFS\cite{Beuzit2019,Claudi2008} and Gemini-S/GPI\cite{Macintosh2008}. The pre-processing applied and the different files provided to the participants are described in Sect.~\ref{sec:instr} and Sect.~\ref{sec:files}, respectively. In each data set we injected 2 to 3 synthetic planetary signals with a realistic spectrum, making a total of $21$ planetary signals to retrieve. The injection procedure, along with its limitations and its application to a SPHERE test data set, is described in Sect.~\ref{sec:inject}. We describe  in Sect.~\ref{sec:ressources} the various resources publicly available on a dedicated Github repository\footnote{\url{https://github.com/exoplanet-imaging-challenge/phase2}}, including a training dataset and Python utilities. At last we illustrate an example injection and retrieval on the training dataset.

\subsection{Pre-reduced data sets}
\label{sec:instr}
The most common observing mode for high-contrast imaging of exoplanets and circumstellar disk consists in carrying out the observations in \emph{pupil-tracking} mode by carefully tuning the derotator of the alt-az mount telescope. This way, the pupil plane is kept in the same orientation, which means that the manifestation of optical aberrations in the image plane remains more or less at the same position. This makes it possible to model and subtract the stellar residuals. Meanwhile, the circumstellar objects rotate in the image field around the target star (centered behind the coronagraphic mask - if any) at a deterministic rate, following the parallactic angles. This so-called \emph{angular diversity} is exploited by the \emph{Angular Differential Imaging} technique\cite{Marois2006}, which consists of three steps: (1) estimating a model of the starlight residuals without circumstellar signals, (2) subtracting this model from each of the frames of the temporal image cube, and (3) combining the subtracted images into a single image therefore containing the subtraction residuals and the circumstellar objects. From this residual map, most post-processing algorithms build a detection map (or probability map, or signal-to-noise ratio map) along with a detection criterion (e.g. a threshold)\cite{Cantalloube2020eidc}.
Because the so-called \emph{ADI sequences} provided by the high-contrast instruments are very common and represents hundreds of available datasets, the second phase of the EIDC is based on such images.

For this second phase we provide eight multispectral temporal image cubes taken with the integral field units (IFU) GPI, installed at the Gemini-South telescope, and SPHERE-IFS, installed at the Very Large Telescope. Both instruments benefit from an exquisite extreme adaptive optics correction, bringing the Strehl ratio up to $90\%$ in the H-band\cite{Fusco2014SAXO,bailey2016gpiao,poyneer2016performance} and an apodized Lyot coronagraph\cite{soummer2005aplc, martinez2009aplc, savransky2014gpiaplc} that makes it possible to reach a raw contrast of $10^{-4}$ at a separation of $500~\mathrm{mas}$ to the star. 
Each data set is taken under different, yet representative, observing conditions, ranging from very good observing conditions of a bright target star to very bad observing conditions of a faint target star. Table~\ref{tab:data} gathers the main parameters of the observation sequence and observing conditions of each of the $8$ data sets.
Additionally, we note that the effective telescope diameter (downstream of the coronagraph) is of $7.57~\mathrm{m}$ for GPI and $7.87~\mathrm{m}$ for SPHERE-IFS and that the plate-scale (the pixel scale of the detector used) is of $14.16~\mathrm{mas/px}$ for GPI and $7.46~\mathrm{mas/px}$ for SPHERE-IFS images. All the images are of odd-number dimension and the center is located on the central pixel.
The centering of the host star is performed as part of the pre-processing.

\begin{table}[!h]
\caption[example] 
   {\label{tab:data} 
Description of the eight data sets provided for the EIDC characterization phase: 
$N_{img}$ is the size of the coronagraphic images; 
$N_{\lambda}$ is the number of channels along the spectral axis; 
$\Delta_{\lambda}$ is the spectral coverage range;
$N_{t}$ is the number of frames along the temporal axis; 
$\Delta_{field}$ is the total field rotation of circumstellar objects. 
The target star is labeled as 'bright' for magnitude in the J-band below 7 and as 'faint' above. 
The last column qualitatively indicates the observing conditions: 'LWE' stands for low-wind effect and 'WDH' stands for wind-driven halo. 
}
\begin{center}
\begin{tabular}{|l| c | c | c | c | c | c | c | c |} 
\hline
Instrument & ID  & $N_{img}$ & $N_{\lambda}$ & $\Delta_{\lambda}$ & $N_{t}$ & $\Delta_{field}$ & Target & Obs. Conditions \\
           &     & $[\mathrm{px}] \times [\mathrm{px}]$& & $[\mathrm{\mu m}]$ & & $[\mathrm{^\circ}]$ &  &  \\
\hline \hline
                    & gpi1 & $169 \times 169$ & $37$ & $1.495 - 1.797$ & $37$ & $14.4$  & bright & very good \\
                    & gpi2 & $169 \times 169$ & $37$ & $1.495 - 1.797$ & $36$ & $156.6$ & bright & medium (WDH) \\
\textbf{GPI}        & gpi3 & $169 \times 169$ & $37$ & $1.495 - 1.797$ & $38$ & $13.3$  & bright & medium \\
                    & gpi4 & $169 \times 169$ & $37$ & $1.495 - 1.797$ & $37$ & $14.2$  & faint & poor (LWE) \\
\hline
                        & sphere1 & $189 \times 189$ & $39$ & $0.957 - 1.329$ & $144$ & $26.8$ & bright & very good \\
                        & sphere2 & $189 \times 189$ & $39$ & $0.957 - 1.329$ & $80$  & $14.4$ & bright & poor (LWE) \\
\textbf{SPHERE-IFS}     & sphere3 & $189 \times 189$ & $39$ & $0.957 - 1.329$ & $64$ & $25.8$ & faint & very good  \\
                        & sphere4 & $189 \times 189$ & $39$ & $0.957 - 1.329$ & $90$ & $144.8$ & bright & poor (WDH) \\
\hline
\end{tabular}
\end{center}
\end{table} 
 
The pre-processing of the raw images delivered by the instruments (i.e. dark subtraction, flat fielding, bad pixel correction, re-centering, cropping, flux normalization and frame selection) were performed by the GPI and SPHERE consortia via the custom-tools they developed: the GPI data cruncher\cite{Perrin2016GPIRP,Maire2010gpi} and the SPHERE data center\cite{Delorme2017sphereDC, Pavlov2008drh}, respectively. The pre-processed data are kindly provided by the two consortia. Before injecting synthetic planetary signals, we homogenized the data (centering, cropping and setting of a parallactic angles direction convention).

\subsection{Files provided to the participants}
\label{sec:files}
The $8$ final complete data sets, containing the injected planetary signals, are provided to the participants on the Zenodo repository in \emph{.fits} file format\cite{Wells1981}. Table~\ref{tab:files} gathers the names of the available files and their content. The header of each file contains general information about the instrumental set-up and observing conditions.

\begin{table}[!h]
\caption[example] 
   {\label{tab:files} 
Description of the five files included for each data set. The `ID' in the filename corresponds to the ones provided in \mbox{Tab.~\ref{tab:data}}. $N_{inj}$ is the number of injected planetary signals in the considered coronagraphic image cube.}
\begin{center}
\begin{tabular}{|l| l | c |} 
\hline
Name & Content & Dimension  \\
\hline \hline
$image\_cube\_ID.fits$ & Multispectral temporal coronagraphic image cube  & 4 \\
$parallactic\_angles\_ID.fits$ & Vectors of parallactic angles and airmass variation & 2 \\
$wavelength\_vect\_ID.fits$ & Vector of the central wavelength of each channel of the IFU & 1 \\
$psf\_cube\_ID.fits$ & Multispectral non-coronagraphic image cube of the target & 3 \\
$first\_guess\_astrometry\_ID.fits$ & First guess position of the injected signals (within $1\lambda/D$) & $N_{inj} \times 2$ \\
\hline
\end{tabular}
\end{center}
\end{table} 

\subsection{Injection procedure}
\label{sec:inject}
The goal of this second phase is not to detect the companions but rather to recover an accurate position and spectrum of the injected companions. In each multispectral image cube, two to three synthetic planetary signals are injected at various locations corresponding to specific starlight residual noise distribution (e.g. at short or large separation, close to low wind effect or wind-driven halo features, etc.). In addition, the signal-to-noise ratio is chosen to assess how it impacts the estimations. At last, because the final goal of planetary spectrum retrieval is to estimate the atmospheric physical parameters of the planet, we injected planets following an emission spectra modeled with a radiative transfer code. In the following we describe the injection procedure that we used in practice. This injection procedure is based on functions included in the Vortex Image Processing (VIP\footnote{https://github.com/vortex-exoplanet/VIP}) and the SPEctral Characterization of directly ImAged Low-mass companions (SPECIAL\footnote{https://github.com/VChristiaens/special}) open-source packages \cite{gonzalez2017vip,Christiaens2022a,Christiaens2021special,Christiaens2022b}.

\subsubsection{Position of the injected signals}
The position of the synthetic planetary signals is of subpixel precision. Fourier-transform based methods are used for shifts and rotations in order to avoid interpolation errors\cite{Larkin1997}. The reference position injected (and to be retrieved) is the signal location in the derotated image. In order to avoid any real astrophysical signal to interfere, we injected the synthetic planetary signals using the opposite parallactic angle variation: it preserves the temporal correlation of the starlight residuals while temporally smearing out any potential true astrophysical signal present in the image cube.

\subsubsection{Spectrum models}
\begin{figure}
    \centering
    \resizebox{\hsize}{!}{\includegraphics{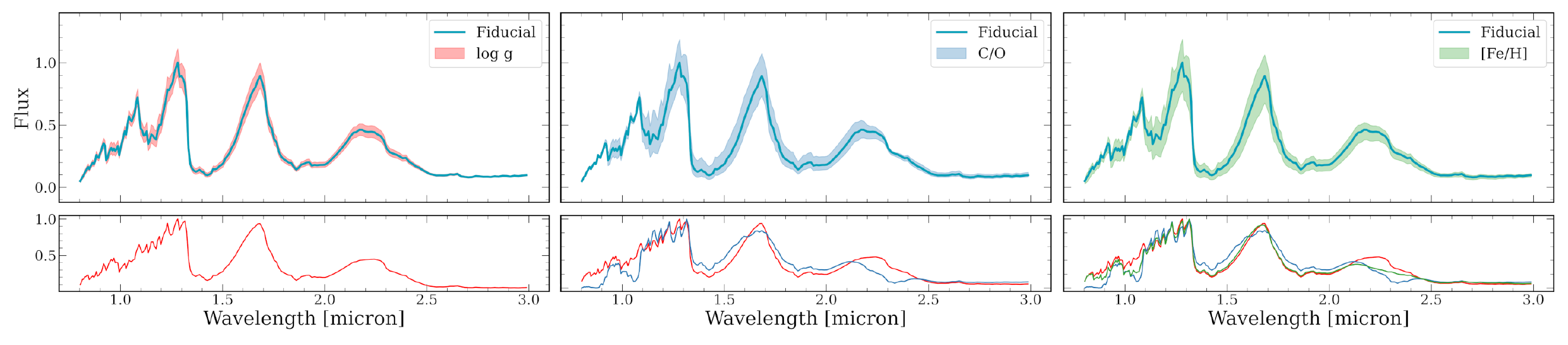}}
    \caption{Typical T-dwarf spectrum obtained with the radiative transfer model PetitRadTrans accross the R- to L-band (blue line) for which key physical parameters are varied. The bottom panels focus on the absolute value of the difference between the highest and lowest part of the parametrized spectra. 
    Left: the surface gravity, log(g), is varied around $3.5^{+0.1}_{-0.1}~\mathrm{cm/s^2}$. 
    Middle: the carbon-oxygen ratio, C/O, is varied around $0.5^{+0.1}_{-0.1}$. 
    Right: the metallicity, [Fe/H], is varied around $0.5^{+0.3}_{-0.2}~\mathrm{dex}$.}
    \label{fig:petitradtrans}
\end{figure}

As illustrated in Fig.~\ref{fig:petitradtrans}, an error in the characterisation stemming from the post-processing has a non-negligible consequence on the physical interpretation of a given planetary emission spectrum. To assess and interpret the effect of the accuracy of the algorithms spectro-photometric estimations on the physical parameters, we injected spectra simulated with the radiative transfer tool petitRADTRANS\cite{Molliere2019petitradtrans,Molliere2020retrieving} that is available in open source\footnote{https://gitlab.com/mauricemolli/petitRADTRANS}. We chose realistic physical parameters (L- or T-dwarf typical spectra, at the transition, cloudy etc.) and sometimes added contribution from circumplanetary material emission by adding a black-body contribution using SPECIAL, and sometimes a bit of noise. The latter also blends the advantages that some algorithms could have because they are taking into account priors on the template spectra for characterization. 
We remind here that the injected spectrum is unknown for the participants since it has to be estimated. 

In order to assess the flux of the host star we performed a 2D-gaussian fit to the non-coronagraphic image cube (file \emph{psf\_cube\_ID.fits}). 
We also took into account the airmass variation during the observing sequence 
such that the injected planet flux in frame $t$ is scaled by $\epsilon_t = \exp(-(X_t - X_{\rm med}))$, where $X_t$ is the airmass value when frame $t$ was acquired, and $X_{\rm med}$ is the median airmass over the observation.
Because the measured spectrum of the star is also affected by both instrumental and atmospheric transmissions, we additionally take these into account for the injection of the synthetic planetary signals. To do so, we use the \emph{NextGen} SED models\cite{Hauschildt1999} for the different observed target stars of the challenge (see Tab.~\ref{tab:stellarspectra} for the assumed stellar parameters), convolved and resampled following the IFUs spectral resolving power and wavelength range, respectively, as implemented in SPECIAL.

\begin{table}[!h]
\caption[example] 
   {\label{tab:stellarspectra} 
Stellar spectra used for each of the 8 data sets. The spectral type, the effective temperature ($T_{eff}$) and surface gravity  ($\log(g)$) are retrieved from the \emph{NextGen} SED models of the 8 observed target stars.}
\begin{center}
\begin{tabular}{|l| c | c | c |} %
\hline
ID & Spectral type & $T_{eff}$ (K) & $\log(g)$ \\
\hline \hline
gpi1 & 	A3V & 8.750 &	4.5 \\
gpi2 &	A0V & 9.600 &	4.5 \\
gpi3 &	F2V & 7.050 &	4.5 \\
gpi4 &	G5V & 5.660 &	4.5 \\
\hline
sphere1 &	B9V & 11.400 & 4.5 \\
sphere2 &	K7V & 4.000  &	4.5 \\
sphere3 &	M0V & 3.750  &	4.5 \\
sphere4 &	A3V & 8.750  & 4.5 \\
\hline
\end{tabular}
\end{center}
\end{table} 

The planetary signals are injected sequentially into the image cube. To choose the signal-to-noise of the injected planetary signals, we run an annular PCA\cite{Soummer2012,Amara2012} (as implemented in VIP) and extract the average contrast over all wavelengths that we define for the injection. Our injection procedure then takes as input: the multispectral coronagraphic image cube in which the injection is made, the normalized non-coronagraphic PSF used as planetary signal model, the parallactic angle and airmass variation through time, the subpixel position for the injection, the planetary spectrum and mean contrast of the injection, as well as the fitted stellar spectrum, the model SED stellar spectrum and finally the instrument spectral resolving power. The planet spectrum is first convolved using the spectral FWHM (inferred from the mean spectral resolution of the instrument), and then resampled to match the IFU wavelength sampling. This is also done with the stellar model SED, before taking the ratio with the measured stellar flux in order to estimate the atmospheric and instrumental transmission to be applied to the planet spectrum. The injected signal $P_{inj}$ for planet $i$ at a given wavelength $\lambda$, position $(x_i,y_i)$ and timestamp $t$ then writes:
\begin{equation}
P_{inj}(x_i,y_i,\lambda, t) = PSF(x_i,y_i,\lambda) \times S_{i}(\lambda) \times \overline{C_{i}} \times T_{atm+instr}(\lambda) \times \epsilon_t,
\label{eq:n}
\end{equation}
were not known to me (but not necessary for me to inject the spectra at given mean contrast levels) - just in case someone asks you
where $PSF(x_i,y_i,\lambda)$ is the non-coronagraphic image of the target star at the wavelength $\lambda$ placed at the position $(x_i,y_i)$, $S_{i}(\lambda)$ is the chosen spectrum of the planet $i$ (simulated by PetitRADTRANS and SPECIAL), $\overline{C_{i}}$ is the mean contrast of the planet $i$ over all the wavelengths probed by the IFU, $T_{atm+instr}$ the transmission of the atmosphere and instrument (i.e.~the measured stellar spectrum via Gaussian fit of the non-coronagraphic spectral images, divided by the convolved+resampled \emph{NextGen} model SED), and $\epsilon_t$ the airmass factor for frame $t$.

\subsubsection{Limitation to the injection procedure}
For the injection procedure, we did not include the following:
\begin{itemize}
    \item No diffraction effect from the coronagraph focal plane mask at close separation;
    \item No smearing effect at large separation (due to long integration time and field of view rotation);
    \item No temporal binning is included;
    \item No additional type of noise is considered on the injection;
    \item In particular, we did not include any other source of flux variability with time (neither intrinsic nor instrumental) apart from the airmass variation;
    \item No jitter (tip-tilt) off-centering the star behind the coronagraph focal plane mask during the exposure: the center of the image is assumed fixed over the whole observation sequence.
\end{itemize}
Note that according to Biller et al., 2021\cite{Biller2021var} the second last point is a correct assumption considering our temporal sampling. For the last point however, it artificially boosts the signal-to-noise ratio compared to reality where it is the major limitations when exploiting angular diversity. 

\subsection{Other available resources}
\label{sec:ressources}
The 8 data sets provided with blind injections are what we call the challenge set, for which the ground-truth remains private and from which the final score of a submission will be computed and displayed in the EvalAI platform leaderboard. Besides this challenge set, we additionally provide one training set consisting of an empty image data set (without injections) from the SPHERE-IFS instrument, along with a procedure to inject synthetic planetary signals. This training set is for the users to estimate hyper-parameters of algorithms when necessary and perform various tests before their final submission on the EvalAI platform. The ID of this empty data set is \emph{sphere0}, a bright target star observed under average conditions, with a total field rotation of $26.8^\circ$ (144 image frames in Y-J band).

On the dedicated Github repository, one can find a toolkit to help participants with processing and submission. Within this toolkit, a folder called \emph{tutorial} contains the tutorial to perform the injections into the training data set (non-blind) following the steps described in Sect.~\ref{sec:inject}. Another folder called \emph{data} contains: (i) the stellar spectra used for the planet injections in each of the $9$ data set provided, and (ii) examples of two planet spectra used for the example injection in the non-blind data set (\emph{sphere0}). This tutorial contains a first part to visualise the input data and a last part to process the data and obtain a quick-look using full-frame PCA-ASDI\cite{Racine1999,Sparks2002} (using, in this case, both temporal and spectral dimensions and 10 principal components to build the model to be subtracted), using its implementation in the VIP package.

\subsection{Illustration with the training dataset (sphere0)}
\label{sec:sphere0}
\begin{figure}
    \centering
    \resizebox{\hsize}{!}{\includegraphics{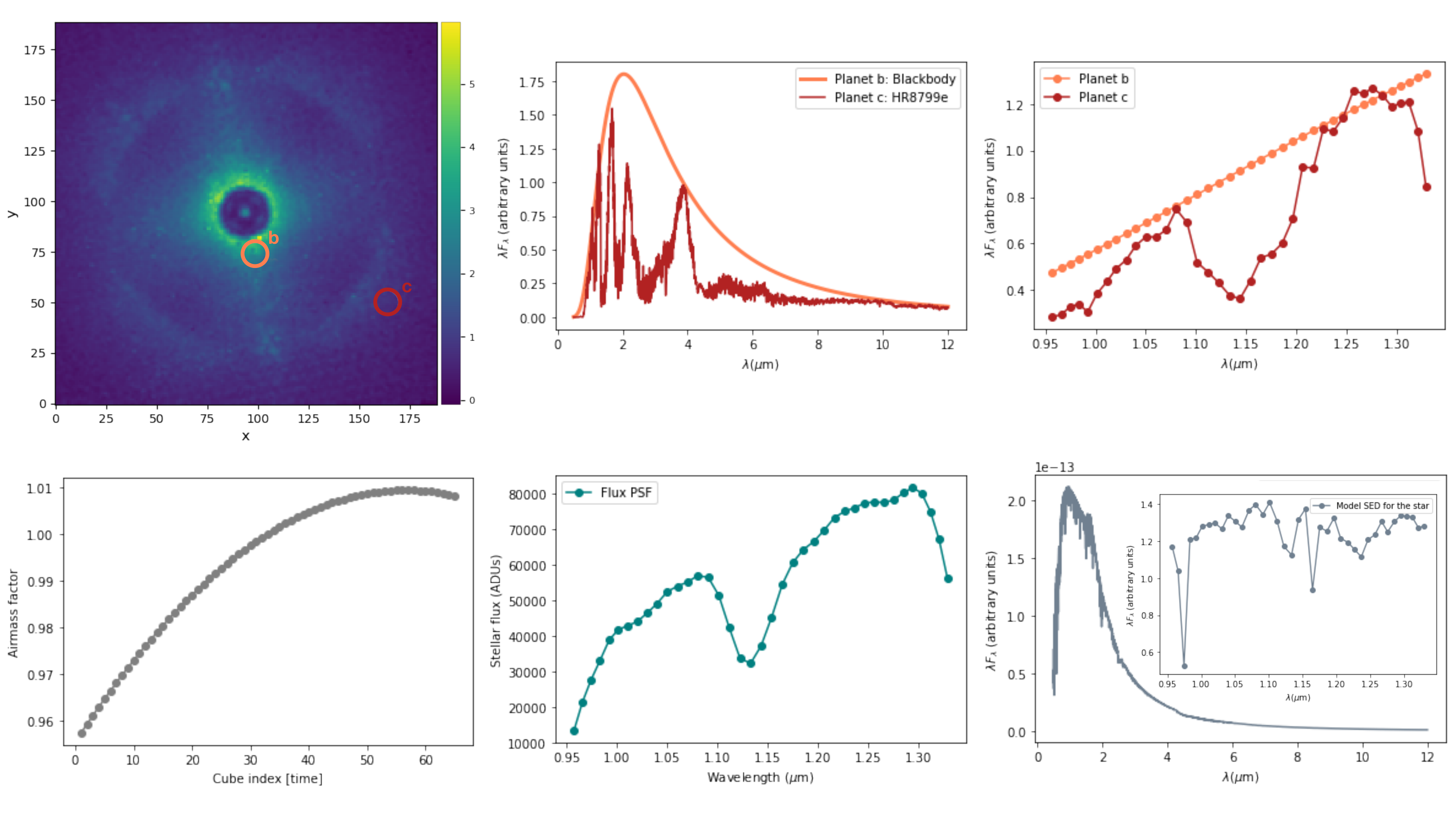}}
    \caption{Example injection procedure for the training data set `sphere0'. Top panel: Median of the image cube at the shortest wavelength with the position of the two injections `b' and `c' indicated by colored circles (left) and their injected spectra of the two planetary signals (middle), sampled at the SPHERE-IFS spectral resolving power (right). 
    Bottom panel: Airmass factor used for the whole observing sequence (left); extracted spectrum of the non-coronagraphic PSF (middle); stellar SED model used for the target star (right), sampled at the SPHERE-IFS spectral resolving power.}
    \label{fig:sph0}
\end{figure}

For the rest of this paper, we consider the example of two injected planetary signals in the \emph{training data set}, dubbed `sphere0'. For this SPHERE-IFS data set, the observing conditions were average (mean seeing $0.99\pm0.22~\mathrm{arcsec}$ and mean coherence time $5.8\pm1.3~\mathrm{ms}$) and the target star rather bright (magnitude $H_{mag}=6.22$ and $V_{mag}\simeq 9.4$), providing with a Strehl ratio of $68\pm10\%$ in H-band. The median image of the cube at the shortest wavelength ($0.957~\mathrm{\mu m}$) is shown in Fig.~\ref{fig:sph0} (top-left panel), along with the location of the two injected planets.

The target was observed for about 1~hour, with an exposure time of $64~\mathrm{s}$, giving $65$ images with a field rotation $\Delta_{field}=22.2^\circ$. The airmass variation during the observing time is shown in Fig.~\ref{fig:sph0} (bottom-left panel). 
The model stellar SED used for calculation of the atmospheric and instrumental transmission (see Eq.~\ref{eq:n}) is a K7V spectral type, with an effective temperature $T_{\rm eff}=4000~\mathrm{K}$ and a surface gravity $log(g)=4.5~\mathrm{cm/m^2}$, whose shape is shown in Fig.~\ref{fig:sph0} (bottom-right panel). The PSF flux extracted from Gaussian fits to the non-coronagraphic multispectral images is shown in Fig.~\ref{fig:sph0} (bottom-central panel). 

The planet `b' signal is injected close to the star at a separation of $155.3~\mathrm{mas}$ (position angle of $167.2^\circ$, measured counter-clockwise from the positive $y$ axis in the image), while planet `c' is injected outside of the AO correction zone, at a separation of $615.9~\mathrm{mas}$ (PA of $122.3^\circ$). 
The model planet spectra considered for the two injections are shown in Fig.~\ref{fig:sph0} (top-central panel). These correspond to a 1800~K black body (blue line) and the best-fit model to planet HR~8799~e (red line)\cite{Nasedkininprep}, for 'b' and 'c' respectively. These spectra are also shown in Fig.~\ref{fig:sph0} after convolution and resampling considering the SPHERE-IFS spectral resolving power and wavelength sampling (top-right panel). The actual spectra injected in the data cube further consider atmospheric+instrumental transmission, and the effect of airmass (see Eq.~\ref{eq:n}). The mean contrast over all the spectral channels chosen for the injections is set to $1.8\times10^{-4}$ and $2.0\times10^{-4}$ for planets b and c, respectively. After a simple full frame ASDI PCA processing, using $10$ principal components to build the model PSF that is subtracted, it corresponds to Signal-to-Noise Ratio values of $5\mathrm{\sigma}$ and $25\mathrm{\sigma}$ in the post-processed image, respectively. In Sect.~\ref{sec:comp}, we compare the results obtained on this data set with two algorithms.

\section{Evaluation and ranking procedure}
\label{sec:eval}
With these 8 data sets (challenge set), there are two tasks to be completed by each participant. The first task is to accurately estimate the astrometry (distance to the star in pixels, using Cartesian coordinates) of all the injected exoplanet signals. The second task is to accurately estimate their spectra (contrast to the star, estimated at each wavelength). Due to the various output that every algorithm may provide, we have decided to focus on distance-based metrics for the  leaderboard of the challenge platform EvalAI. This simple metric is detailed in this section.
 
\subsection{First task: Astrometry}
\label{sec:astro}
For this first task, the EvalAI leaderboard displays the distance (in the sense of the L2-norm, euclidian distance) between the estimated value and the ground truth value. The final result is the mean of this distance for all the 21 injections:
\begin{equation}
\begin{split}
D_{astrometry} &= \frac{1}{ N_{inj}} \; . \; \sum_{i} | \hat{p_i} - p_i^{GT}|_2 \\
&= \frac{1}{ N_{inj}} \; . \; \sum_{i} \sqrt{( \hat{x_i} - x_i^{GT})^2 + (\hat{y_i} - y_i^{GT})^2},
\end{split}
\label{eq-astro}
\end{equation}
with $\hat{p_i}$ the estimated position for the injection $i$, $p_i^{GT}$ its corresponding ground truth, and $N_{inj}$ the total number of injections. In this challenge the position of the planetary signals are to be provided as the number of pixels, in Cartesian coordinates, with respect to the star at the center of the image frame. 
The leaderboard displays this value, the closer to 0 the better.

Note that participants who do not find a given injected planet can provide input values of '-1' or 'NaN' in the result \emph{.fits} files. In that case, the leaderboard will display the value obtained with Eq.~(\ref{eq-astro}) after replacing the corresponding $\hat{p}_i$ with a position at $1 \lambda/D$ from the ground-truth position (for the longest wavelength). In the final paper we will however weight this non-detection to penalize the final ranking accordingly.

\subsection{Second task: Spectro-photometry}
\label{sec:photo}
For this second task, the EvalAI leaderboard displays the distance (in the sense of the L1-norm) between the estimated value and the ground truth value for each spectral channel separately. We then average the results over all the spectral channels. The final result is the mean of this distance for all the 21 injections:
\begin{equation}
  D_{photometry} = \frac{1}{N_{inj}} \; . \; \frac{1}{N_{\lambda}} \; . \; \sum_{i} \sum_\lambda \frac{| \hat{c}_{\lambda,i} - c^{GT}_{\lambda,i} |_1}{c^{GT}_{\lambda,i}},  \label{eq-photo}
\end{equation}
with $\hat{c}_{\lambda}$ the estimated contrast for the injection $i$ at wavelength $\lambda$, $c^{GT}_{\lambda}$ its corresponding ground truth, and $N_{inj}$ the total number of injections and $N_{\lambda}$ the total number of spectral channels of the IFU. The absolute value of the difference is normalized by the ground-truth contrast in order to consider the same weights for same relative differences in inferred contrasts (be it in shallow or high contrast parts of the spectra) in the final metric. 
The leaderboard displays this value, the closer to 0 the better.

Note that participants who do not find a given injected planet should retrieve the flux using the first guess astrometry provided in the data set. If they still cannot extract a value for the photometry, they can provide input values of '-1' or 'NaN' in the result \emph{.fits} files. In that case, the leaderboard will display the value of Eq.~(\ref{eq-photo}) after replacing the corresponding $\hat{c}_{\lambda,i}$ values by 0. In the final paper we will however weight this non-extraction to penalize the final ranking accordingly.

\subsection{Limitation to the evaluation procedure}
\label{sec:limites}
This distance-based metric does not take into account two important aspects: (i) the uncertainties on the estimations and (ii) the difficulty of each retrieval per injection. We made this decision because some algorithms (e.g. algorithms based on supervised machine learning from the artificial intelligence field) do not provide any uncertainties on the estimations. In addition the uncertainties depend on an underlying probability density function that could either be estimated and provided by the participants or assumed Gaussian (whereas we know this assumption is often not respected\cite{Pairet2019}). Taking into account this uncertainty is therefore not trivial. 


For the second point, we decided to ignore it in a first step as most algorithms should naturally adjust to it. 
For the next communication giving the complete results of this second phase during the year 2023, we plan to use more evolved assessment methods taking into account those two aspects by e.g. weighting the grade with the signal-to-noise ratio and uncertainties (if any).

\subsection{Submission guidelines}
\label{sec:limites}
To participate, users must submit on the EvalAI platform ZIP files containing their predictions. To participate in both tasks, two ZIP files must be uploaded under the following name: \emph{username\_astrometry.zip} and \emph{username\_photometry.zip}. Each of these ZIP files will contain as many Multiple Extension FITS (MEF) files as data cubes to evaluate, i.e eight files per task. The name of each MEF .fits file must follow the format of \emph{astrometry\_ID.fits} and \emph{photometry\_ID.fits} (with `ID' provided in Table \ref{tab:data}). Each MEF fits file should contain three extensions, one for the estimated parameters, one for the uncertainties on these estimated parameters (optional) and one for posterior distributions over the estimated parameters (optional). 
The second and third dimensions (uncertainties and posterior respectively) are optional input that will be used for further analyses in order to better understand the performance of a given algorithm.

We remind here that if the participant does not succeed in detecting one of the injected planetary signals, they can enter values of '-1' or 'NaN' in the final result \emph{.fits} files. For the leaderboard ranking these values are replaced with appropriate quantities (see previous sections), but for the final discussion in a future publication, a finer criterion will be defined to penalize the final score accordingly.

In the Github repository dedicated to the second phase of the exoplanet imaging data challenge, a folder called \emph{eidc2} contains the code that the EvalAI platform runs to compute the leaderboard ranking, as well as the code computing the metrics chosen for the leaderboard ranking. In the \emph{tutorial} folder of the EIDC phase 2 Github repository, there is a tutorial to create a multi-extension \emph{.fits} file.

\section{Results using two methods: PCA-NEGFC and ANDROMEDA}
\label{sec:preres}
In this section, we show the results obtained with two techniques widely used in the community to infer the position and contrast of the two planets injected in the training data set (Sect.~\ref{sec:sphere0}). We compare the estimated astrometry and contrasts to the ground truth, and apply the metrics defined in Secs.~\ref{sec:astro} and \ref{sec:astro} to evaluate the score achieved by each method.
In our previous paper\cite{Cantalloube2020eidc}, we reviewed the main detection algorithms that currently exist and sorted them in four main families. Similarly, available techniques in the community for characterization of planetary signals in high-contrast images can be sorted into two broad families: (a) the fake-companion injection methods, 
and (b) inverse-problem methods.

The fake companion injection methods are based on the injection of synthetic planetary signals using the non-coronagraphic PSF of the star so as to make a forward model. Different implementations have been proposed. The imprint of the companion on the calculated principal components can be considered to build a forward model which captures the effects of over- and self-subtraction and is then directly fitted to the stellar PSF subtracted image (as e.g. in KLIP-FM\cite{Pueyo2016}). Alternatively negative synthetic planetary signals can be injected in the image cube before applying the differential imaging technique, with the flux and position of the negative injection varied until the residuals in the post-processed images are minimized at the location of the detection (i.e. NEGFC\cite{marois2010negfc,lagrange2010negfc}). In either cases, the minimization can be performed via statistical sampling methods, e.g. by running a Markov Chain Monte Carlo (MCMC) fitting algorithm\cite{foreman2013emcee} in order to estimate the parameters and their uncertainties\cite{bottom2015mcmc}. 

The inverse problem approach consists in building a model of the planetary signal (after differential imaging or without differential imaging) by using the normalized PSF (non-coronagraphic or unsaturated image of the star) and estimating the contrast of any planetary signal present in the field of view via a maximum likelihood estimation\cite{Mugnier2009}. Algorithms such as ANDROMEDA\cite{cantalloube2015andro}, FMMF\cite{ruffio2017fmmf}, PACO\cite{flasseur2018paco}, and RSM-FM\cite{dahlqvist2021rsmfm} are part of this family.

\subsection{Results with PCA-NEGFC}
\label{sec:pcanegfc}


We used the simplex algorithm of the NEGFC implementation in VIP in order to find the optimal position and contrast relative to the star of the companions injected in the training data set\cite{Wertz2017,Christiaens2021special}. More specifically we proceeded as follows. We obtained a first guess estimate of the injected position of the companions in a PCA-ASDI processed image (10 principal components used). We then inferred the optimal number of principal components which maximized the SNR at that position in each individual spectral channel, using the \verb|pca_annulus| function of VIP -- this function along with these optimal number of principal components are then used for all subsequent steps. Next, we considered the top 5 channels in terms of SNR for the extraction of the astrometry in a combined NEGFC involving 7 free parameters: 2 for the position and 5 for the individual fluxes in each of these spectral channels. Once the astrometry was inferred, we fixed the position of the companion and ran the NEGFC simplex algorithm to derive the flux of the companion individually in each of the 39 channels.

The astrometric uncertainties on `b' and `c' were estimated with the injection of 360 fake companions at the same radius as `b' and `c', and with the same estimated flux as `b' and `c', respectively. The injections were performed individually, 1 deg apart from each other, and the same parameters were used for the retrieval of their position and contrast using the NEGFC simplex algorithm as for `b' and `c'. We fitted a Gaussian to the distribution of deviations between the retrieved positions and the known injected positions, to infer the 1 sigma uncertainties on the radial separation and azimuth\cite{Wertz2017}. A similar method was adopted to infer the 1-sigma uncertainty on the contrast of `b' and `c' in each channel.

\begin{figure}
    \centering
    \resizebox{\hsize}{!}{\includegraphics{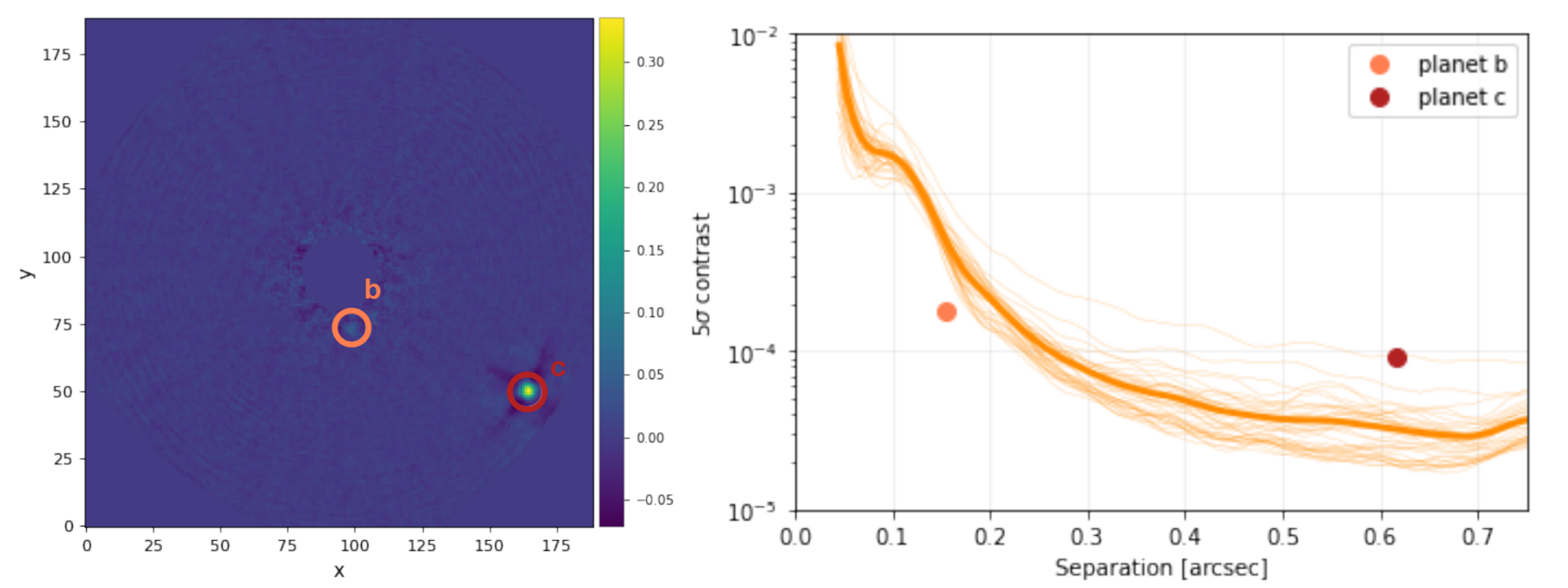}}
    \caption{Results of the 'sphere0' data set processed with principal component analysis. Left: Final residual map obtained with PCA-ASDI using 10 principal components. Right: Detection limit as 5-sigma contrast curves obtained with PCA-ADI. The 39 thin lines are the contrast curve for each channel of the spectro-imager. The thick line is the mean of all the thin lines. The two dots show the radial position and mean contrast of the two injected planetary signals.}
    \label{fig:detmappca}
\end{figure}

Figure~\ref{fig:detmappca} shows the final PCA-ASDI image (left) and the 5-sigma contrast curves obtained with PCA-ADI (10 PCs) in each spectral channel (right). The characterization results obtained PCA-NEGFC are shown in Fig~\ref{fig:astro} and Fig~\ref{fig:phot}, where they are compared to the ground truth astrometry and contrast values, respectively (orange).

\subsection{Results with ANDROMEDA}
\label{sec:andro}
ANDROMEDA is based on an inverse problem approach\cite{Mugnier2009}: it includes a model of the expected planetary signal (as the normalized non-coronagraphic image of the star) and fits its position and flux. A first step is to perform a pairwise subtraction of a couple of images for which the field has rotated enough to not subtract the planetary signal but that are temporally close in time so that the starlight residuals are still correlated enough. This first step allows on the one hand to whiten the residual starlight and on the other hand to create a specific signature if a planetary companion is present. The key idea of ANDROMEDA is to estimate the flux and position of any planetary signal in the field of view by using a maximum likelihood estimation, under the hypothesis that the residuals after the pairwise subtraction is white and Gaussian\cite{cantalloube2015andro}. 

In order to retrieve the astrometry of the two injected companions, we ran ANDROMEDA with the default parameters in SADI mode. This mode first performs a spectral differential imaging after rescaling a reference image at a different wavelength (set to the first spectral channel in our case) before running the classical ADI-based ANDROMEDA. ANDROMEDA provides one signal-to-noise ratio (SNR) map per spectral channel. The final detection map is obtained by summing the squared SNR maps and then taking its square root before normalizing it to obtain an equivalent SNR map. The astrometry is estimated directly on this latter detection map by performing a 2D-Gaussian fit on the signals that are above a threshold set to $5\sigma$. The associated uncertainties are driven by the errors of the 2D-Gaussian fit performed on the planet detection signal. The spectro-photometry is estimated at the position of the planetary signals extracted on the final detection map, showing the highest SNR. The flux is read for each spectral channel directly on each map of the maximum likelihood estimation of the flux (classical ADI-based ANDROMEDA). The associated uncertainties are therefore driven by the signal-to-noise ratio of the planetary signal. 

\begin{figure}
    \centering
    \resizebox{\hsize}{!}{\includegraphics{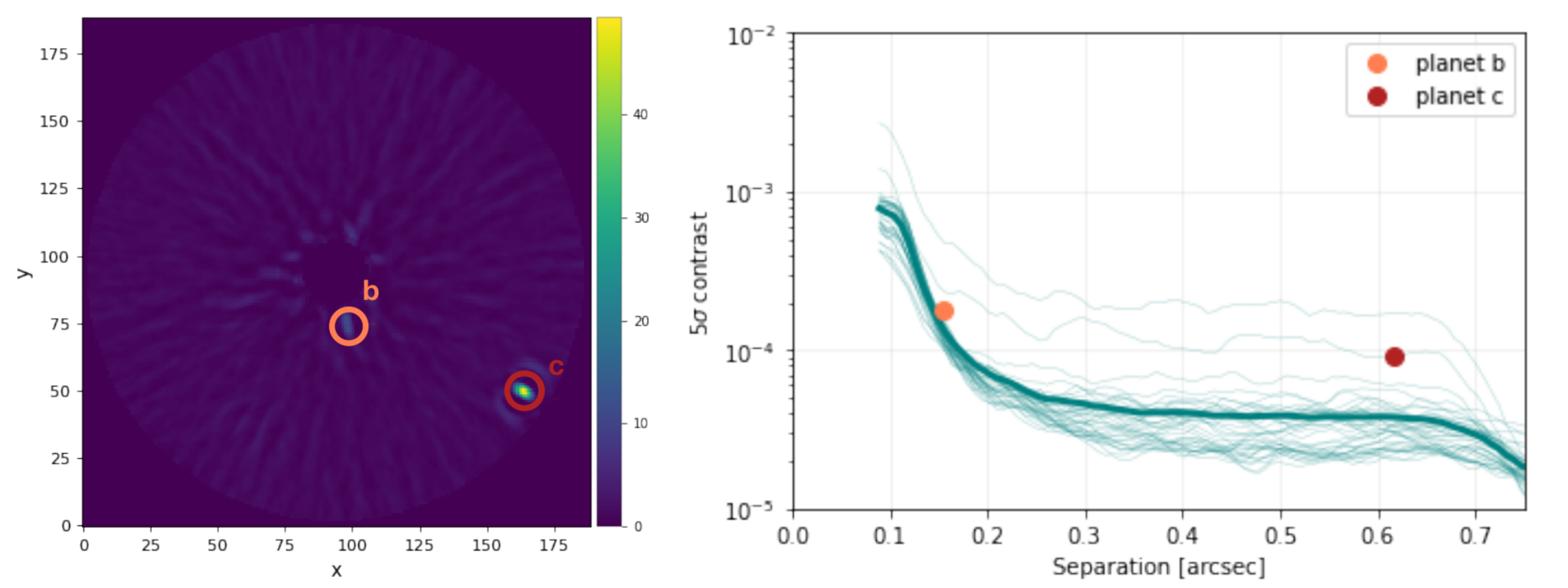}}
    \caption{Results of the 'sphere0' data set processed using ANDROMEDA performing first a spectral image difference (after wavelength rescaling) and then an angular pairwise difference (SADI mode). Left: Final signal-to-noise ratio map on which the detection and astrometry estimation is performed. Right: Corresponding detection limit as 5-sigma contrast curves. The 39 thin lines are the contrast curve for each channel of the spectro-imager. The thick line is the mean of all the thin lines. The two dots show the radial position and mean contrast of the two injected planetary signals.}
    \label{fig:detmapandro}
\end{figure}

Figure~\ref{fig:detmapandro} shows the final residual map (left) and the associated 5-sigma contrast curves (right). The SNR for the planet 'b' is of $10$, and for the planet 'c' is $48$. The characterization results obtained with ANDROMEDA are shown in Fig~\ref{fig:astro} and Fig~\ref{fig:phot} with respect to the ground truth (blue).

\subsection{Comparison}
\label{sec:comp}
We applied both PCA-NEGFC and ANDROMEDA on the training data of the data challenge (sphere0) with the injection described in Sect.~\ref{sec:sphere0}. The results of the metrics are presented in Tab.~\ref{tab:res}. Note that when the planetary signal has not been detected, the first guess provided in the data can be used to extract the flux. 

\begin{table}[!h]
\caption[example] 
   {\label{tab:res} 
Results of the astrometry and spectrophotometry metrics described in Sect.~\ref{sec:eval} obtained with PCA-NEGFC and ANDROMEDA for the two injections.}
\begin{center}
\begin{tabular}{|l| c | c |c|| c | c |c|} 
\hline
\multirow{2}{*}{\textbf{Method}} & \multicolumn{3}{c||}{PCA-NEGFC} & \multicolumn{3}{c|}{ANDROMEDA} \\ 
\cline{2-7}
& planet b & planet c & all & planet b & planet c & all \\
\hline \hline
Astrometry &  0.37 &  0.03 & 0.20 & 1.95 & 0.06 & 1.01\\
\hline
Spectro-photometry & $16.98$ & $1.03$ & 9.00 & 35.67 & 2.29 & 18.99\\
\hline
\end{tabular}
\end{center}
\end{table} 

The first injected planet 'b' is very close to the star. Current algorithms, based on ADI and/or SDI are limited in this region due to the poor amount of diversity and the larger starlight residuals. This explains why, as a general rule planet b has a higher metrics in both astrometry and spectrophotometry compared to planet c that is very bright and far from the star, in a region with diversity and less affected by starlight residuals. 

\subsubsection*{Astrometry estimation}
\begin{figure}
    \centering
    \includegraphics[scale=0.6]{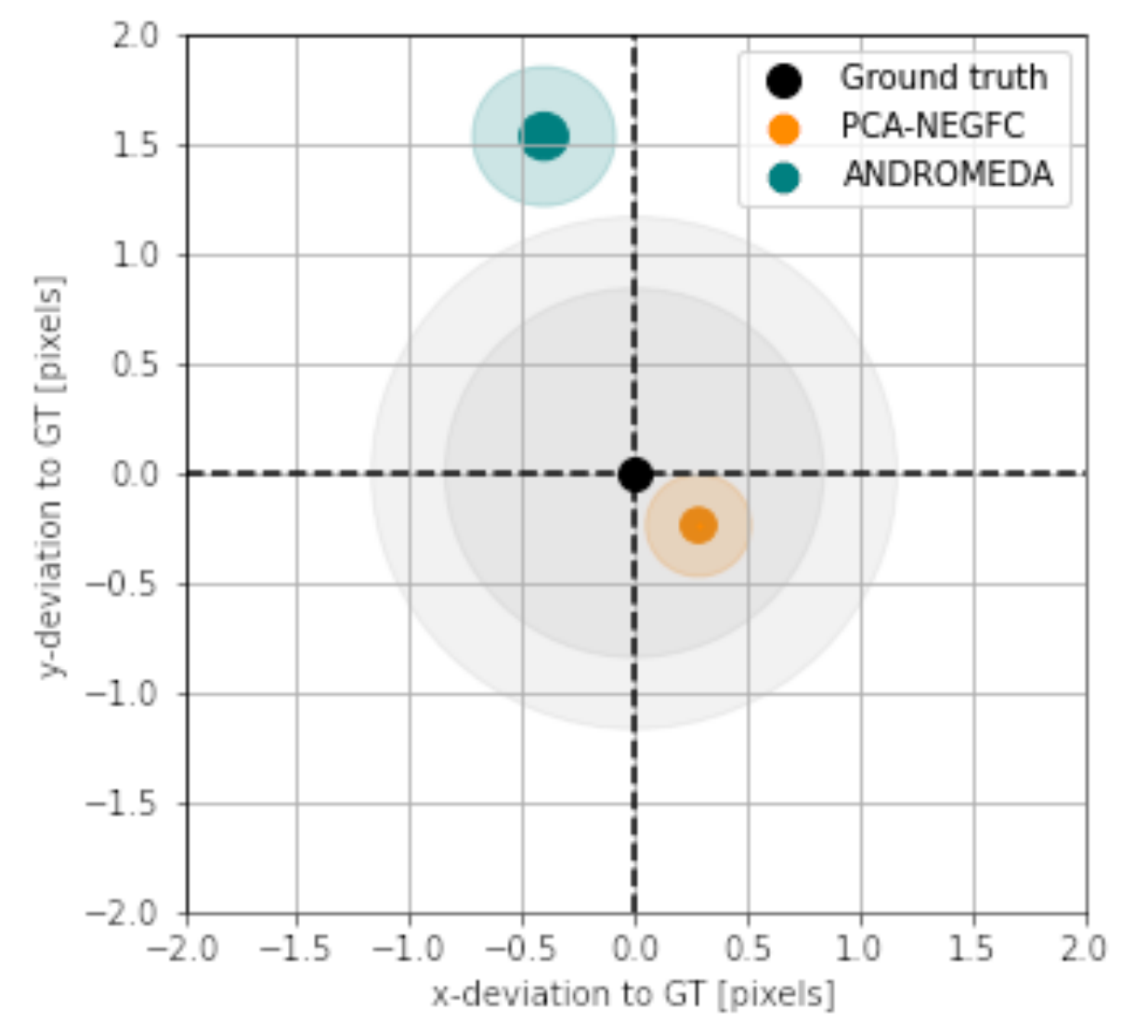}
    \includegraphics[scale=0.6]{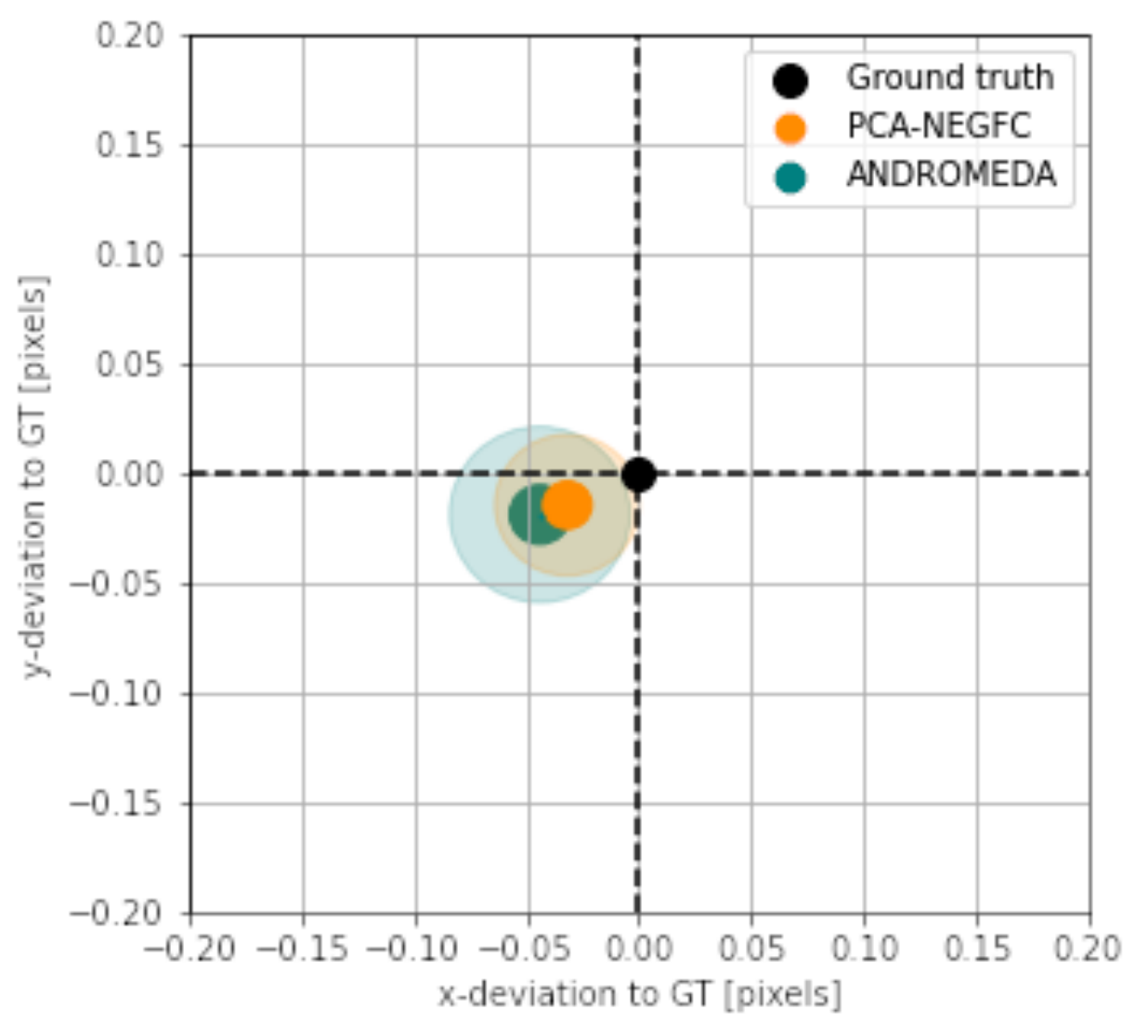}
    \caption{Astrometry estimations of the injected planets using PCA-NEGFC (orange) and ANDROMEDA (blue), relative to the ground truth (black) in Cartesian coordinates. The size of the symbols corresponds to the 1-sigma uncertainty (averaged in x- and y- direction) and the shaded area to the 3-sigma uncertainty. The two centered grey shaded areas represent the size of a half resolution element ($0.5\lambda/D$) for the shortest and largest wavelengths of the spectro-imager. 
    Left: Planet 'b' ($155~\mathrm{mas}$). Right: Planet 'c' ($616~\mathrm{mas}$).}
    \label{fig:astro}
\end{figure}

The results of the astrometry estimations analysis are shown in Fig.~\ref{fig:astro} for both planets (left and right). Note that the grid is 10 times smaller for the right panel (planet c) than the left panel (planet b).

In addition to being located at short separation, planet b lies on a bright speckle, which significantly affects the estimation by ANDROMEDA. Indeed the SNR spot at the location of planet b is quite elongated radially (see Fig.~\ref{fig:detmapandro}, left), which biases the estimation towards a closer estimated separation (by about $7-8~\mathrm{mas}$, which is one pixel). The NEGFC technique appears to be less affected by the bright speckle, possibly owing to  the astrometry estimate being made only in the 5 channels with maximum SNR. However, for neither method does the ground truth fall within the 3-sigma uncertainties provided by the NEGFC method. 

As for planet c, both methods provide estimations within $\sim$0.05 pixel from the ground truth, although it is worth mentioning that the associated uncertainties appear slightly underestimated in both cases.

\subsubsection*{Spectrophotometry estimation}
\begin{figure}[t!]
    \centering
    \includegraphics[scale=0.5]{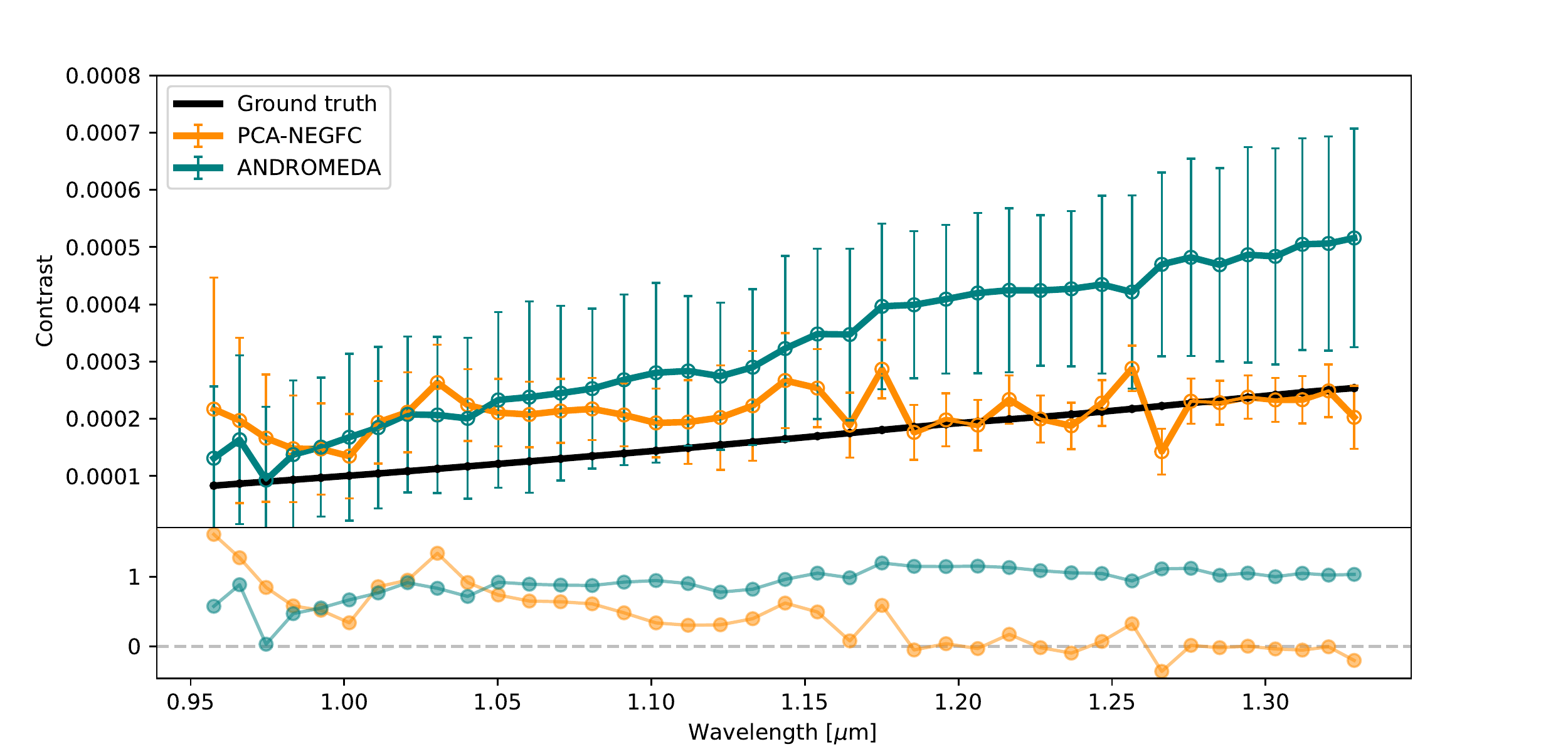}\\
    \vspace{0.26cm}
    \includegraphics[scale=0.5]{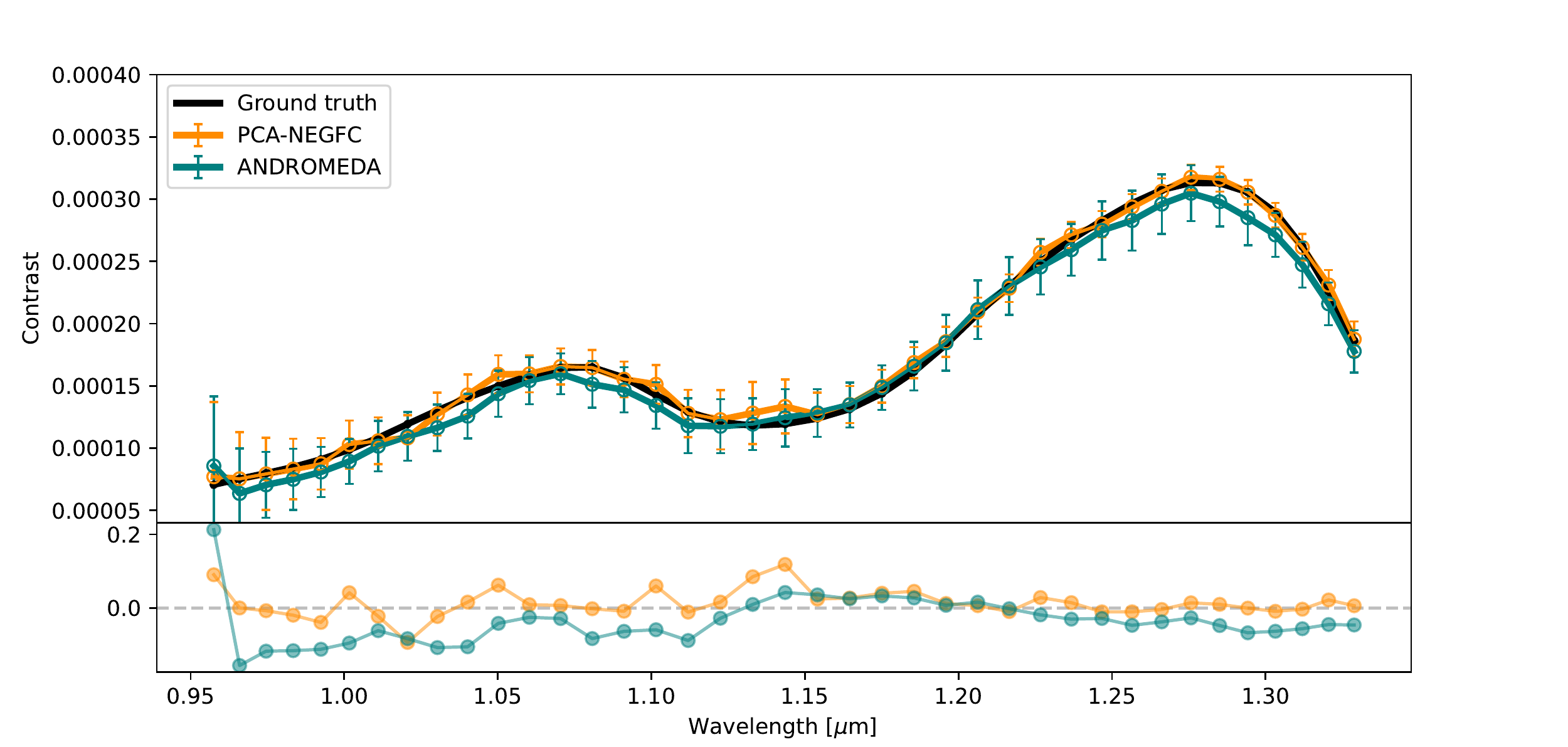}
    \caption{Spectro-photometry estimations of the injected planets using PCA-NEGFC (orange line) and ANDROMEDA (blue line), compared to the ground truth (black line), as well as the residuals normalized by the ground truth (bottom panel). The error bars displayed correspond to the 3-sigma uncertainties on the photometry estimation. 
    Top: Planet 'b' (Blackbody spectrum). 
    Bottom: Planet 'c' (HR8799e-like spectrum).}
    \label{fig:phot}
\end{figure}

The results of the photometry estimations analysis are shown in Fig.~\ref{fig:phot} for both planets (top and bottom). Note that the cuts are not the same for the two panels.

For planet b, we note that ANDROMEDA provides the correct trend but seems to be always slightly higher than the ground truth, in particular in the redder spectral channels (i.e. the residuals are flat with wavelength). This can again be explained by the presence of the bright speckle that lies below the injection and that crosses the injection location at larger wavelength. On the contrary, PCA-NEGFC slightly overestimates the bluer spectral channels, while extracting very accurate values at longer wavelengths (i.e. the residuals are decreasing with wavelength). PCA-NEGFC provide a rather flat spectrum. This can be understood by the residual minimization procedure performed by NEGFC that is more aggressive towards residual speckles. 
However, we can see that even though both methods give results within 5-sigma from the ground truth, this is still suboptimal as many points are not within 3-sigma (for both methods), demonstrating that there is still room for improvement.

For planet c, PCA-NEGFC provides an almost perfect estimate of the spectrum. ANDROMEDA provides slightly worst results at bluer wavelengths. This is probably because the starlight residuals distribution is more intense and shaped in this part of the spectrum and ANDROMEDA heavily relies on the assumption that the subtraction residuals are Gaussian and white, which we now know is not a perfect assumption\cite{Pairet2019}.
Both methods have larger uncertainties at short wavelength due to the inherent shape of the spectrum. 

In conclusion, we can already see that this comparison, based on the same dataset and the same injected signals, provides a lot of information about the fundamental limitations of the two methods used here, which are based on very different concepts. Having more data, for various instruments, under various observing conditions and for injections of varying complexity (in terms of contrast and localization) will hopefully provide us with valuable information on the functioning of the algorithms and their limitations. This work can therefore offer ways to improve the various methods that exist today and open avenues for the development of new methods.

\section{Conclusion \& Perspectives}
Thanks to this second phase of the challenge, we hope to understand better the differences we often observe in the estimation of flux and position of the planet candidates, as a function of the chosen post-processing technique along with its user-parameters. This full exercise will provide us with an insight on the global error made due to the intrinsic limitations of all the algorithms and subsequently on how it affects the retrieval of physical parameters. We therefore highly encourage the high-contrast community to participate in the challenge.

After the deadline, the working group will work on a complete analysis of the results, including all of the submitted algorithm outputs. Instead of using the simple ranking described in the current paper, we will define a more complex metric that will notably account for the difficulty of the retrieval for each injection and each data set by, for instance, weighting our figure of merit accordingly. 

For the next phase of the \emph{Exoplanet Imaging Data Challenge}, we intend to focus on circumstellar disk imaging. Indeed, new dedicated methods were recently developed (e.g. MAYONNAISE\cite{pairet2021mayonnaise}, NMF-DI\cite{ren2020dataimputation}, REXPACO\cite{flasseur2021rexpaco}) and classical methods adapted (e.g. iterative PCA\cite{pairet2018iterative}, 
NMF\cite{ren2018nmf}) to this purpose. The goal of this third phase is to detect the circumstellar signal but also to restore accurately its total intensity distribution for structures such as rings, spirals and shadows. Within this phase, we intend to make multi-reference differential imaging (mRDI\cite{lafreniere2009rdi, ruane2017rdi}) technique possibles by using SPHERE and GPI data, for which a large collection of archival data is available (making sure the data set of the challenge is not included in the library). 

As a fourth phase, we envision to focus on High Resolution Spectroscopy (HRS), with instruments such as VLT/SINFONI\cite{eisenhauer2003sinfoni}, VLT/MUSE\cite{bacon2010muse}, VLT/CRIRES+\cite{follert2014crires+} and Keck/NIRCSPEC\cite{mclean1998nirspec}, as techniques based on molecular mapping are currently under development\cite{snellen2014mm,rameau2021mm}. Another focus would be to simulate ELT high contrast images to see which post-processing technique is the most adapted and get ready with the image processing tools before first light\cite{houlle2021harmoni}. As a conclusion, the \emph{Exoplanet Imaging Data Challenge} community initiative has various tasks to carry out in order to support the community of exoplanetary system observers and image processing developers to make the most out of the data delivered by the latest generation of instruments.


\acknowledgments 
This project has received funding from the European Research Council (ERC) under the European Union's Horizon 2020 research and innovation programme (grant agreement No 819155), and from the Wallonia-Brussels Federation (grant for Concerted Research Actions). 
This work is partly based on data products produced at the SPHERE Data Centre hosted at OSUG/IPAG, Grenoble. SPHERE is an instrument designed and built by a consortium consisting of IPAG (Grenoble, France), MPIA (Heidelberg, Germany), LAM (Marseille, France), LESIA (Paris, France), Laboratoire Lagrange (Nice, France), INAF–Osservatorio di Padova (Italy), Observatoire de Genève (Switzerland), ETH Zurich (Switzerland), NOVA (Netherlands), ONERA (France) and ASTRON (Netherlands) in collaboration with ESO.

\bibliography{report} 
\bibliographystyle{spiebib} 

\end{document}